\documentclass[twocolumn,showpacs,preprintnumbers,amsmath,amssymb]{revtex4}

\begin{document}

\title{Quantum bit string sealing}
\author{Guang-Ping He}
\affiliation{Department of Physics and Advanced Research Center,
Zhongshan University, Guangzhou 510275, China}

\begin{abstract}
Though it was proven that secure quantum sealing of a single
classical bit is impossible in principle, here we propose an
unconditionally secure quantum sealing protocol which seals a
classical bit string. Any reader can obtain each bit of the sealed
string with an arbitrarily small error rate, while reading the
string is detectable. The protocol is simple and easy to be
implemented. The possibility of using this protocol to seal a
single bit in practical is also discussed.
\end{abstract}

\maketitle

\newpage

\section{Introduction}

Quantum sealing is a young member of quantum cryptography. But classical
sealing has entered our everyday life for centuries. For example, people
sometimes put important letters or documents inside an envelope, and seal
the envelope by melting wax over the cover flap. Additionally, the wax can
be impressed with an image to indicate authenticity, such as a family crest.
If the wax was broken or the image looks unmatched, then someone may have
opened the envelope and read the document. Thus it provides a method to
check whether the secret document remains secure or not. Obviously it is
useful to expand the idea of classical sealing to the digital world.
However, just as\ other classical cryptographic protocols, no classical
sealing protocol can be unconditionally secure. This is because there is no
no-cloning theorem of classical data to prevent a cheater from copying and
reading all the data without being detected.

On the other hand, quantum cryptography has made significant progress in the
last two decades. Many quantum protocols, e.g. the conjugate coding\cite
{Wiesner} and the well-known quantum key distribution\cite{BB84,Ekert91,B92}%
, surpassed their classical counterparts as their security is rested on the
basic laws of quantum mechanics and can be unconditionally secure. Therefore
it is natural to ask whether the quantum no-cloning theorem\cite{no-cloning}
can make secure quantum sealing possible. Bechmann-Pasquinucci\cite{sealing}
proposed the first quantum sealing protocol in 2003 which seals a single
classical bit with a three-qubit state, shortly followed by Chau\cite{seal q}
with a protocol which seals quantum data with quantum error correcting code.
In 2005, Singh and Srikanth\cite{Srikanth} extended the idea of Ref.\cite
{sealing}\ into a many-qubit majority voting scheme, and associated it with
secret sharing to improve the security. These protocols can be divided into
two types. The first two are perfect quantum sealing in the sense that a
reader can obtain the sealed data with certainty. The last one is imperfect
for the reader cannot do so accurately.

However, as pointed out by the authors themselves, the protocols
in Refs.\cite{sealing, seal q} are insecure against collective
measurements. More general, it was proven\cite{impossibility} that
secure perfect quantum sealing is impossible in principle. With
collective measurements, the reader can always read the data
without being detected. Very recently, the general models of
imperfect quantum sealing were studied by He\cite{He} and
Chau\cite {Chau} with different approaches, showing that they are
also insecure for sealing a single classical bit. The reader can
always cheat successfully with a non-trivial probability.
Consequently, the majority voting scheme in Ref.\cite{Srikanth} is
insecure alone, if it is not associated with secret sharing. That
is, its security has to fall back on that of the secret sharing
scheme. The existence of these three no-go proofs seems to put an
end to the development of quantum sealing.

This is not true. Though secure quantum sealing of a single
classical bit is impossible, it will be shown in this paper that
quantum sealing of a classical bit string can be unconditionally
secure. At the first glance, this result seems odd since classical
reasoning suggests that secure bit string sealing implies secure
single bit sealing and hence conflicts with the above no-go
proofs. But as pointed out by Kent\cite{QBSC}, reductions and
relations between classical cryptographic tasks need not
necessarily apply to their quantum equivalents. Similar situation
happened before in quantum cryptography. Though the possibility of
unconditionally secure quantum bit commitment was excluded by the
Mayers-Lo-Chau no-go theorem\cite {Mayers,LC}, Kent found that
secure quantum bit string commitment protocol exists\cite{QBSC}.
In this paper, a secure quantum bit string sealing protocol will
be proposed. It achieves the following goal: each bit of the
string can be obtained by the reader with an arbitrarily small
error rate, while reading the string can be detected except with
an exponentially small probability. The significance of the
protocol lies in three major aspects. (1) In practical, it is
obvious that sealing a long bit string is much more useful than
sealing a single bit. (2) As shown by Gordon Worley III\cite
{Gordon}, secure quantum sealing, if exists, has wide applications
in many different fields, e.g. loose bit commitment, binary
semaphores, eavesdropping detection and protective packaging.
Therefore our protocol can re-open new venues for these
applications that once closed by the no-go proofs. (3) The
protocol is very simple to be implemented with the techniques
available nowadays. In additional, the protocol also makes it
possible to implement ``computationally'' secure quantum bit
sealing in practical.

\section{The Protocol}

Quantum bit string sealing can be summarized as the following two-party
cryptographic problem. A sender Alice encodes an $n$-bit string with quantum
states. Any reader Bob can obtain the string from these states, while
reading the string should be detectable. Clearly quantum bit sealing can be
viewed as the special case where $n=1$.

Consider the ideal case without transmission error. Let $\Theta $\ ($%
0<\Theta \ll \pi /4$) and$\ \alpha $\ ($0<\alpha <1/2$) be two fixed
constants. We propose the following quantum bit string sealing protocol:

\bigskip

\textit{Sealing}: To seal a classical $n$-bit string $b=b_{1}b_{2}...b_{n}$ (%
$b_{i}\in \{0,1\}$), Alice randomly chooses $\theta _{i}$ ($-\Theta
/n^{\alpha }\leqslant \theta _{i}\leqslant \Theta /n^{\alpha }$) and encodes
each bit $b_{i}$ in a qubit state $\left| \psi _{i}\right\rangle =\cos
\theta _{i}\left| b_{i}\right\rangle +\sin \theta _{i}\left| \bar{b}%
_{i}\right\rangle $. She makes these $n$ qubits publicly accessible to the
reader, while keeping all $\theta _{i}$ ($i=1,...,n$) secret.

\textit{Reading}: When Bob wants to read the string $b$, he simply
measures each qubit in the computational basis $\{\left|
0\right\rangle ,\left| 1\right\rangle \}$, and denotes the outcome
as $\left| b_{i}^{\prime }\right\rangle $. He takes the string
$b^{\prime }=b_{1}^{\prime }b_{2}^{\prime }...b_{n}^{\prime }$ as
$b$.

\textit{Checking}: At any time, Alice can check whether the sealed string $b$
has been read by trying to project the $i$-th qubit into $\cos \theta
_{i}\left| b_{i}\right\rangle +\sin \theta _{i}\left| \bar{b}%
_{i}\right\rangle $. If all the $n$ qubits can be projected successfully,
she concludes that the string $b$ is still unread. Otherwise if any of the
qubits fails, she knows that $b$ is read.

\bigskip

Obviously this new protocol achieves the following goal: each bit sealed by
Alice can be read successfully by Bob, except with a probability not greater
than $\varepsilon \equiv \sin ^{2}(\Theta /n^{\alpha })$. Thus by increasing
$n$, the reading error rate $\varepsilon $\ can be made arbitrarily small.

In the next section it will be proven that the protocol also guarantees
that: if the string was read, it will be detected by the checking process
except with an exponentially small probability.

\section{Proof of Security}

When the reader Bob measures the $i$-th qubit in the basis $\{\left|
0\right\rangle ,\left| 1\right\rangle \}$, the qubit collapses to $\left|
b_{i}^{\prime }\right\rangle $. We may even assume Bob to be malevolent,
that he replaces each qubit with another quantum state $\left| \psi
_{i}^{\prime }\right\rangle $ after reading it, so that his chance to pass
the checking without being detected might be increased. Due to the
no-cloning theorem of quantum states\cite{no-cloning}, Bob cannot determine
and copy the state $\left| \psi _{i}\right\rangle $ exactly since he does
not know $\theta _{i}$. He cannot even be sure whether $b_{i}^{\prime }=b_{i}
$ or not after he read $\left| \psi _{i}\right\rangle $. Thus he has to pick
another $\theta _{i}^{\prime }$ ($-\Theta /n^{\alpha }\leqslant \theta
_{i}^{\prime }\leqslant \Theta /n^{\alpha }$) himself and prepares the fake
state as $\left| \psi _{i}^{\prime }\right\rangle =\cos \theta _{i}^{\prime
}\left| b_{i}^{\prime }\right\rangle +\sin \theta _{i}^{\prime }\left| \bar{b%
}_{i}^{\prime }\right\rangle $. Since the case $b_{i}^{\prime }=b_{i}$ (or $%
b_{i}^{\prime }=\bar{b}_{i}$) will occur with the probability $\cos
^{2}\theta _{i}$ (or $\sin ^{2}\theta _{i}$), the fake state $\left| \psi
_{i}^{\prime }\right\rangle $ can be projected to $\left| \psi
_{i}\right\rangle =\cos \theta _{i}\left| b_{i}\right\rangle +\sin \theta
_{i}\left| \bar{b}_{i}\right\rangle $ successfully in the checking process
with the probability
\begin{equation}
p_{i}=\cos ^{2}\theta _{i}\cos ^{2}(\theta _{i}-\theta _{i}^{\prime })+\sin
^{2}\theta _{i}\sin ^{2}(\theta _{i}+\theta _{i}^{\prime }).
\end{equation}
When $\theta _{i}$ is evenly distributed among the range $[-\Theta
/n^{\alpha },\Theta /n^{\alpha }]$, the average of $p_{i}$\ is
\begin{equation}
\bar{p}_{i}=\frac{1}{2\Theta /n^{\alpha }}\int\nolimits_{-\Theta /n^{\alpha
}}^{\Theta /n^{\alpha }}p_{i}d\theta _{i}.
\end{equation}
Its maximum can be reached when $\theta _{i}^{\prime }=0$. That is, it is
better for Bob not to fake the state, but simply leaves the $i$-th qubit as
it is after measuring it. In this case
\begin{equation}
p_{i}=1-\frac{1}{2}\sin ^{2}2\theta _{i}.
\end{equation}
Therefore the total probability for Bob to read $k=\beta n$ ($0\leqslant
\beta \leqslant 1$) bits without being detected is
\begin{equation}
P=\prod\limits_{i=i_{1}}^{i_{k}}(1-\frac{1}{2}\sin ^{2}2\theta _{i}),
\end{equation}
which drops exponentially as $k\rightarrow n$, and vanishes when $%
n\rightarrow \infty $ as long as $0<\alpha <1/2$.

However, in the more general case Bob may not read the string with the
measurement suggested by the protocol. He may not even want to learn each $%
b_{i}$ individually, but tries to perform collective measurement on the
whole system $\Psi =\psi _{1}\psi _{2}...\psi _{n}$ so that he can obtain
some global properties of the string (e.g. parity, weight etc.). In this
case, let $H$ denotes the $2^{n}$\ dimensional Hilbert space where $\Psi $
lives in. Suppose that $\Psi $ finally collapses into a subspace $V$ after
Bob performs certain POVMs. Let $\{v\}$ and $m$ be the computational basis
and the dimensionality of $V$ respectively. Then no matter how the details
of Bob's cheating strategy could be, the amount of information Bob obtained
is bound by
\begin{equation}
k=\log _{2}2^{n}-\log _{2}m.
\end{equation}
Meanwhile, the final state of $\Psi $ is
\begin{equation}
\left| \Psi ^{\prime }\right\rangle =\frac{1}{N}\sum\limits_{v\in V}\left|
v\right\rangle \left\langle v\right| \left. \Psi \right\rangle ,
\end{equation}
where the normalization constant
\begin{equation}
N=(\sum\limits_{v\in V}\left| \left\langle v\right| \left. \Psi
\right\rangle \right| ^{2})^{1/2}.
\end{equation}
Again, it can be shown that it is better for Bob not to fake the state. Then
$\left| \Psi ^{\prime }\right\rangle $ can be projected to the initial state
$\left| \Psi \right\rangle =\left| \psi _{1}\right\rangle \left| \psi
_{2}\right\rangle ...\left| \psi _{n}\right\rangle $\ successfully in the
checking process with the probability
\begin{equation}
P=\left| \left\langle \Psi \right| \left. \Psi ^{\prime }\right\rangle
\right| ^{2}=\sum\limits_{v\in V}\left| \left\langle v\right| \left. \Psi
\right\rangle \right| ^{2}.
\end{equation}
Since $\left| \psi _{i}\right\rangle =\cos \theta _{i}\left|
b_{i}\right\rangle +\sin \theta _{i}\left| \bar{b}_{i}\right\rangle $\ and $%
\{v\}$ is the computational basis, for any $v$\ we have
\begin{equation}
\left| \left\langle v\right| \left. \Psi \right\rangle \right| ^{2}\leqslant
\prod\limits_{i=1}^{n}\cos ^{2}\theta _{i}.
\end{equation}
Therefore the total probability for Bob to obtain $k$ bits of information
without being detected is
\begin{equation}
P\leqslant m\cdot \prod\limits_{i=1}^{n}\cos ^{2}\theta
_{i}=2^{-k}\prod\limits_{i=1}^{n}2\cos ^{2}\theta _{i}.
\end{equation}
which also drops exponentially as $k\rightarrow n$, and vanishes when $%
n\rightarrow \infty $ as long as $0<\alpha <1/2$.

As a result, no matter Bob reads the string with individual or
collective measurements, the probability for him to avoid from
being detected will always be exponentially small as the amount of
information he obtained increases. Thus the protocol is
unconditionally secure.

\section{Discussions}

\subsection{Relationship with the no-go proofs}

The existence of this secure quantum bit string sealing protocol does not
conflict with the no-go proofs of quantum single bit sealing\cite
{impossibility,He,Chau}. In fact, our quantum bit string sealing protocol
can be viewed as the assembly of $n$ imperfect quantum single bit sealing
process. From the security proof in the above section, we can see that if
Bob reads only few bits, the disturbance on the quantum states is small that
it is almost undetectable. In this sense, the sealing of each single bit of
the string is insecure. Also, it is insecure to use the global properties of
the string (e.g. parity, weight etc.) to implement single bit sealing. But
if Bob reads a large number of bits, the small disturbance on every single
qubit will be piled up together so that the detecting probability will
increase dramatically. Hence the sealing of the whole string can be secure.

\subsection{The protocol is an imperfect sealing one}

The sealed string can only be read with a non-zero error rate $\varepsilon
\equiv \sin ^{2}(\Theta /n^{\alpha })$. We cannot associate the protocol
with classical error-correcting codes or any other method to make the string
perfectly retrievable. This is because the above security proof is based on
the fact that Bob has no pre-knowledge on the string and the quantum states.
If he is provided with a certain classical error-correcting code or anything
relevant with the sealed string, he may have other methods to construct his
collective measurements so that the security proof may not be valid any
more. In fact, it is trivial to show that the no-go proof of perfect quantum
bit sealing\cite{impossibility} can be generalized to the case of perfect
quantum bit string sealing. If the whole string becomes perfectly
retrievable, then each bit of the string is perfectly retrievable too. From
the proof in Ref.\cite{impossibility} we can see that if Bob reads every
single bit with collective measurements, the disturbance will rigorously
equal to zero. Thus the total detecting probability will not be piled up but
still equal to zero. Therefore, though the error rate $\varepsilon $\ can be
made arbitrarily small, we cannot expect to find methods to make it
completely vanished.

On the other hand, as pointed out in Ref.\cite{He}, the no-go proof of
imperfect quantum single bit sealing does not cover the case of string
sealing. More rigorously, in our protocol by expanding the state $\left|
\Psi \right\rangle $ in the computational basis of the global Hilbert space $%
H$, we can see that $\left| \Psi \right\rangle $ covers all eigenvectors of $%
H$.\ That is, the ``sub''-space supported by $\left| \Psi
\right\rangle $ is exactly the space $H$\ itself. Therefore the
spaces supported by different states which encode different
strings completely overlap with each other. No measurement can
distinguish them apart without disturbing the states seriously.
Thus the cheating strategy in the no-go proofs of imperfect
quantum single bit sealing\cite{He,Chau} does not apply here.

\subsection{Implementability}

Our protocol can be executed as long as Alice has the probability to prepare
each single qubit in a pure state, while Bob can perform individual
measurement. No entanglement or collective measurement required. Therefore
the protocol can be demonstrated and verified with the techniques available
nowadays. Of course for practical uses, storing quantum states for a long
period of time is still a technical challenge today. But this is a problem
which all quantum sealing protocols have to face. Our protocol may already
be one of the simplest in all possible quantum bit string sealing protocols.

\section{Sealing a single bit in practical}

Though secure sealing of a single bit is impossible in principle, if a
protocol can be found in which reading the bit dishonestly is much more
difficult than doing so honestly, it will still be valuable in practical.
The no-go proof of imperfect quantum bit sealing\cite{He} leaves a clue on
how to construct such a protocol. As pointed out in that reference, quantum
sealing protocol generally contains the following feature: Bob knows an
operation $P$ and two sets $G_{0}$, $G_{1}$, such that if he applies $P$ on
the quantum system that seals the bit\ and the outcome is $g\in G_{0}$ (or $%
g\in G_{1}$), he should take the value of the sealed bit as $0$ (or $1$).
Though in principle we cannot exclude this feature from the protocol
(otherwise the sealed bit becomes irretrievable), we can keep the dishonest
reader from knowing $P$, $G_{0}$ and $G_{1}$ too easily. The method is:
Alice can seal the description of $P$, $G_{0}$ and $G_{1}$ with the quantum
string sealing protocol. If Bob wants to decode the sealed bit correctly, he
should read this sealed string first. Thus the status of the sealed bit can
be checked by detecting whether the sealed string has been read.

For example, Alice first encodes the following sentence into a classical
binary bit string
\begin{eqnarray*}
&&``\mathit{Measure\ the\ last\ two\ qubits\ in\ the\ basis\ } \\
&&\{\cos 15%
%TCIMACRO{\UNICODE[m]{0xb0}}%
%BeginExpansion
{{}^\circ}%
%EndExpansion
\left| 0\right\rangle +\sin 15%
%TCIMACRO{\UNICODE[m]{0xb0}}%
%BeginExpansion
{{}^\circ}%
%EndExpansion
\left| 1\right\rangle ,-\sin 15%
%TCIMACRO{\UNICODE[m]{0xb0}}%
%BeginExpansion
{{}^\circ}%
%EndExpansion
\left| 0\right\rangle +\cos 15%
%TCIMACRO{\UNICODE[m]{0xb0}}%
%BeginExpansion
{{}^\circ}%
%EndExpansion
\left| 1\right\rangle \}\mathit{\ } \\
&&\mathit{and\ you\ will\ know\ the\ value\ of\ the\ sealed\ bit\ from\ } \\
&&\mathit{their\ parity.\ Other\ qubits\ following\ this\ sentence\ } \\
&&\mathit{are\ all\ dummy\ qubits.\ You\ can\ simply\ leave\ them\ } \\
&&\mathit{alone.}"
\end{eqnarray*}
Then she seals it with our quantum string sealing protocol, and provides Bob
the qubits encoding this sentence, followed by a large number of qubits
where only the last two are actually useful.

However, this bit sealing method is still insecure in principle. This is
because any given classical $n$-bit string can be decoded into one sentence
only, and the meaning of the sentence will reveal the value of the sealed
bit unambiguously. As long as a dishonest Bob knows the length $n$\ of the
sealed string, he can study all the $2^{n}$ possible classical $n$-bit
strings, decode them into sentences, and divide these sentences into $G_{0}$
and $G_{1}$ (of course there will also be tons of meaningless sentences. Bob
can simply leave them alone). Then as described in Ref.\cite{He}, he needs
not to know the content of the sealed sentence exactly. He simply constructs
a proper collective measurement to determine whether the sentence belongs to
$G_{0}$ or $G_{1}$. Thus he will know the sealed bit from the $n$ qubits
without disturbing them too much.

But if $n$ is sufficiently large, the number of possible sentences will be
enormous. There could be sentences as simple as
\[
``\mathit{It\ is\ }0\mathit{.\ Ignore\ the\ rest\ qubits.}"
\]
But there are also sentences like
\begin{eqnarray*}
&&``\mathit{Decode\ the\ bits\ following\ this\ sentence\ as\ a\ } \\
&&\mathit{bitmap\ image\ and\ you\ will\ find\ clues\ to\ the\ } \\
&&\mathit{value\ of\ the\ sealed\ bit}"
\end{eqnarray*}
or
\begin{eqnarray*}
&&``\mathit{Go\ to\ the\ main\ library.\ Find\ the\ book\ on\ the\ } \\
&&\mathit{top-left\ of\ the\ last\ shelf.\ Turn\ to\ the\ last\ page,\ } \\
&&\mathit{and\ count\ how\ many\ times\ the\ letter\ K\ occurs\ } \\
&&\mathit{in\ the\ 3rd\ line......}"
\end{eqnarray*}
or even
\begin{eqnarray*}
&&``\mathit{Dig\ my\ backyard\ until\ you\ find\ water.\ Count\ } \\
&&\mathit{how\ many\ feet\ you\ digged.\ Then\ divide\ it\ by\ the\ } \\
&&\mathit{height\ of\ the\ tree\ }\mathit{in\ the\ north\ corner......}"
\end{eqnarray*}
In this case, even if a dishonest Bob has the technique to perform
collective measurements on the $n$-qubit system, in practical it is nearly
impossible to check all these sentences and find out the bit value they are
corresponding to. On the other hand, an honest Bob needs not to worry about
this. He can simply read the sealed string honestly and then follows the
instruction to decode the sealed bit. In this sense, such sealing can be
viewed as a kind of ``computationally'' secure quantum single bit sealing in
practical.

I would like to thank Helle Bechmann-Pasquinucci for useful discussions.


\begin{thebibliography}{99}
\bibitem{Wiesner}  S. Wiesner, SIGACT News, \textbf{15}, 78 (1983).

\bibitem{BB84}  C. H. Bennett, G. Brassard, in \textit{Proceedings of IEEE
International Conference on Computers, Systems, and Signal Processing},
Bangalore, India, pp.175 (IEEE, New York, 1984).

\bibitem{Ekert91}  A. K. Ekert, Phys. Rev. Lett. \textbf{67}, 661 (1991).

\bibitem{B92}  C. H. Bennett, Phys. Rev. Lett. \textbf{68}, 3121 (1992).

\bibitem{no-cloning}  W. K. Wootters and W. H. Zurek, Nature \textbf{299},
802 (1982).

\bibitem{sealing}  H. Bechmann-Pasquinucci, Int. J. Quant. Inform. \textbf{1}%
, 217 (2003).

\bibitem{seal q}  H. F. Chau, quant-ph/0308146.

\bibitem{Srikanth}  S. K. Singh, R. Srikanth, Physica Scripta \textbf{71},
433 (2005).

\bibitem{impossibility}  H. Bechmann-Pasquinucci, G. M. D'Ariano, C.
Macchiavello, quant-ph/0501073. To be published in Int. J. Quant. Inform.
(2005).

\bibitem{He}  G. P. He, quant-ph/0502179. To be published in Phys. Rev. A
(2005).

\bibitem{Chau}  H. F. Chau, quant-ph/0503031.

\bibitem{QBSC}  A.Kent, Phys. Rev. Lett. \textbf{90}, 237901 (2003).

\bibitem{Mayers}  D. Mayers, Phys. Rev. Lett. \textbf{78}, 3414 (1997).

\bibitem{LC}  H. -K. Lo and H. F. Chau, Phys. Rev. Lett.\textbf{78}, 3410
(1997).

\bibitem{Gordon}  G. Gordon Worley III, quant-ph/0504207. SPIE paper 5815-25
at Quantum Information and Computation III, SPIE Defense \& Security
Symposium 2005.
\end{thebibliography}
\end{document}